\documentclass[12pt]{article}  
\usepackage{graphics}
\usepackage{a4,epsfig}
\usepackage{lineno, amssymb}

\begin{document}

\pagestyle{plain}
\begin{center}


{\bf {\Large On Stimulated Radiation of Black Holes}}

\vspace{1cm}
{Jan Ridky}

\vspace{0.5cm}
{\small{\em FZU AV CR, v.v.i., Na Slovance 2, 182 21 Praha 8, Czech Republic}}
\vspace{0.5cm}
\begin{abstract}
\noindent
The Unruh's thermal state in the vicinity of the event horizon of the black hole provides conditions where impinging particles can radiate other particles. The subsequent decays may eventually lead to observable radiation of photons and neutrinos induced even by massive particles with gravitational interaction only. The hadronic particles will induce $\sim 30$ MeV  gamma radiation from $\pi^{0}$ decays.

\end{abstract}
\end{center}
\vspace{0.5cm}


\noindent
The Schwarzschild radius and the event horizon are the notions of classical physics and general relativity. The quantum aspects of black holes appeared first in the papers of Hawking \cite{Hawking1,Hawking2} which changed radically our understanding of the phenomena connected with them and introduced their thermal radiation. Together with the work of Unruh \cite{Unruh1,Unruh2} this conceptual framework provides a basis for better understanding of phenomena occurring close to the event horizons of black holes. It was shown in \cite{Unruh1} that the fiducial observer located outside the Schwarzschild radius will see the vacuum near the event horizon as a many-particle thermal ensemble. In standard interpretation the free falling observer will perceive this as Heisenberg fluctuations while the accelerated observer will detect these thermal particles as also shown in \cite{Unruh1}. Thus orbiting particle with {\em{perimelasma}} \footnote{The terminology is not yet settled, see {\em http://en.wikipedia.org/wiki/Perihelion}} gradually shrinking to the value of Schwarzschild radius will ultimately interact with the thermal particles. 

The quantum phenomena taking place around the event horizon of a black hole pose a formidable problem of quantization in a curved space. Actually \cite{Hawking1} tackles the problem by describing asymptotic states of vacuum in flat space before and after collapse of a star. The relation between accelerated and non-accelerated systems is studied in \cite{Unruh1}. At first the quantization of massive scalar field in Rindler coordinate system is compared to that in Minkowski space as in the case of large black hole the use of flat space-time is locally valid approximation of the horizon. The quantization in Rindler coordinates based on \cite{Fulling} leads to the result inequivalent with the standard quantization in Minkowski space, namely the respective vacua differ. The vacuum in Minkowski space can be expressed as thermal many-particle ensemble of the Unruh-Fulling-Rindler (UFR) state. The procedure is examined also in the Schwarzschild metric, in this case the problem with definition of positive frequency waves has to be addressed, but the result remains valid. The temperature of the UFR state increases to infinity as the Rindler variable 
\begin{center}
$\rho = \int^{r}_{2M} (1-\frac{2M}{r'})^{-\frac{1}{2}}dr'$
\end{center}
approaches to zero, i.e. the observer approaches the Schwarzschild radius $R = 2M$, where 
$M$ is the mass of the black hole and we put $1=G=\hbar=c=k \;(Bolzmann's \; constant)$. This holds for any black hole regardless of its mass while the temperature of the Hawking radiation changes with its mass as $8\pi/M$. Indeed, a black hole can radiate a particle of mas $m$ if it drags down another particle from the UFR state beyond the event horizon so that the conservation laws are satisfied. The intensity of gravitational field close to the Schwarzschild radius behaves as $m/M$. Thus the probability of a particle with mass $m$ to be radiated decreases with the mass $M$ of the black hole. The black hole radiation is entirely determined by the surface gravity $\kappa$ introduced in \cite{BCH} and it plays the role of temperature, in the case of a black hole with zero angular momentum $\kappa \sim 1/M$.

\begin{figure}[htb]
  \begin{center}
       \includegraphics*[width=13cm,height=7cm]{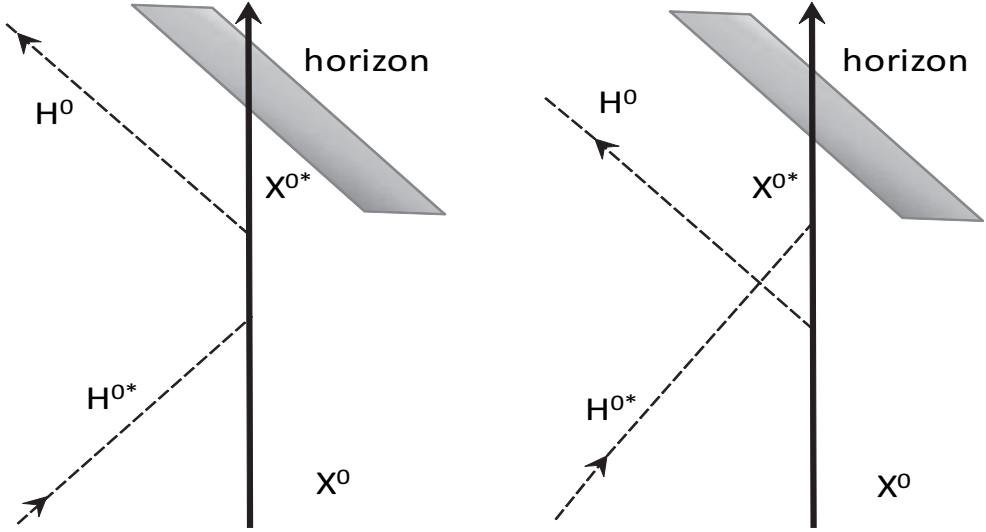}
    \caption{Scattering of $X^{0}$ with Higgs boson $H^{0*}$ in the UFR state.}
    \label{fig:CDM-H}
  \end{center}
\end{figure}

As mentioned above we can observe the UFR vacuum state. This is examined in detail in \cite{Unruh1}. It is shown that accelerated detector - testing particle or field, interacts with particles of the vacuum state, provided these two fields have nonzero coupling. Interesting consequences arise if we implement this abstract example to the Standard Model or its extension. The UFR state is then occupied by all SM particle species and the orbiting particle can interact with them. The {\em stretched horizon} is introduced as the horizon several Planck lengths away from the mathematical horizon in \cite{{SusskindLindesay}} and it is shown that it has properties of an electric conductor. Thus the dynamics of a charged particle in the vicinity of the stretched horizon might be quite complex. The most simple case will be that of a massive neutral particle $X^{0}$ with gravitational interaction only. To conciliate this assumption with extension of SM we can consider this particle as an example of cold dark matter. Then besides gravitation this particle interacts with Higgs bosons $H^{0}$ provided it is sufficiently close to the horizon, i.e. the UFR state has a temperature corresponding to the occurrence of Higgs bosons. With the increasing proximity to the stretched horizon and with increasing temperature the symmetries are gradually restored. Thus the particles will be generally off mass shell. The most simple process is then
$X^{0}H^{0*} \rightarrow H^{0}X^{0*}$.
The proximity of the event horizon and the environment of the UFR state allow that the mass $m(X^{0*}) = m(X^{0})-\Delta$ and $\Delta > 0$, i.e. $X^{0}$ gets off mass shell. The situation is sketched in Fig.~\ref{fig:CDM-H}. 
As all participating particles are massive we can use locally flat space-time. Of course we cannot describe the gravitational interactions of $X^{0}$ by means of quantum field theory but we can try to estimate their size from classical physics. The definition of the Schwarzschild radius is equivalent to the condition that at this radius the potential energy of a particle in the gravitational field of a black hole equals to one half of its energy. Hence this gives an upper bound on the size of gravitational effects. After passing the event horizon the particle forms a bounded state with the black hole and its mass is decreased by the mass defect. Thus the radiated Higgs boson can gain the energy compared to the one it had in the UFR state on the expense of the initial mass of $X^{0}$. In the case of particles coming in radial direction the mass defect will be completely dissipated inside the limits of event horizon while in the other limiting case of particles moving in tangential direction toward the stretched horizon, the subsequent decay of radiated $H^{0}$ and the following cascade of decays may produce photons and neutrinos observable far from the event horizon. This all of course under the condition that $\Delta$ is sufficiently large compared to the $m(H^{0})-m(H^{0*})$ mass difference. This mechanism can lead to observation of matter dominated cold dark matter consisting of massive particles with gravitational interaction only, without annihilations in space. Such case is usually set aside, e.g. in \cite{PDM}, as a difficult one to treat.

The recent discovery of neutrino oscillations implies that neutrinos are massive and they will be not radiated as the Hawking radiation until the black hole temperature will increase sufficiently. Thus the largest parts of massive black holes will be radiated in the form of photons. The eventual radiated neutrinos would thus originate from other possible sources and part them also from the above described stimulated radiation. In the hypothetically possible case of the lightest neutrino still massless the Hawking radiation would produce only this one species of neutrinos while in neutrinos produced by stimulated radiation all flavors will be represented and their energy spectrum will be also different.

In a similar way the accreting protons will produce $\pi^{\pm}$ and $\pi^{0}$. Provided the movement of $\pi^{0}$ with respect to the black hole is not relativistic some of $\gamma$ originating from $\pi^{0}$ decays will emerge at the energy of $\sim 30$ MeV. They will be further subjected to the additional red shift due to the rotation of the accreting material and the movement of a given black hole.

The intensity of the radiation stimulated by black holes will be mainly determined by the amount of impinging particles and it will depend on the surroundings of the black hole. Thus it is reasonable to expect that the more massive black holes with larger surfaces of event horizon will radiate more. This is in contrast with Hawking thermal radiation where the more massive black holes are cooler.

The author is grateful to his colleagues Dalibor Nosek, Rupert Leitner, Radomir Smida and Petr Travnicek for comments and discussion. The work was supported by the project of the Academy of Sciences of the Czech Republic AV0Z10100502 and the project LA08016 of the Ministry of Education of the Czech Republic.


\begin{thebibliography}{99}

\bibitem{Hawking1} S. Hawking, Nature {\bf{248}} (1974), 30.

\bibitem{Hawking2} S. Hawking, Commun. math. Phys. {\bf{43}} (1975), 199.

\bibitem{Unruh1} W. G. Unruh, Phys. Rev. {\bf{D14}} (1976), 870.

\bibitem{Unruh2} W. G. Unruh, Phys. Rev. {\bf{D15}} (1977), 365.


\bibitem{Fulling} S. Fulling, Phys. Rev. {\bf{D7}} (1973), 2850.

\bibitem{BCH} J. M. Bardeen, B. Carter, S. W. Hawking, Commun. math. Phys. {\bf{31}} (1973), 161.

\bibitem{SusskindLindesay} L. Susskind, J. Lindesay, {\em{AN INTRODUCTION TO BLACK HOLES, INFORMATION and the STRING THEORY REVOLUTION}}, (World Scientific Publishing Co. Pte. Ltd., 2005)

\bibitem{PDM} G. Bertone, D. Hooper, J. Silk, Phys. Rep. {\bf{405}} (1973), 279.

\end{thebibliography}
\end{document}